# Angular compliance, not amplitude, governs the dielectric response of $BaTiO_3$

Ryotaro Sahashi[1], Po-Yen Chen[1], and Teruyasu Mizoguchi[1,2,*]

[1]Department of Materials Engineering, The University of Tokyo, Tokyo, Japan
[2]Institute of Industrial Science, The University of Tokyo, Tokyo, Japan
[*]Contact author: teru@iis.u-tokyo.ac.jp

**ABSTRACT**. Soft-mode theory specifies a frequency, not what a local polar unit does. What real-space motion underlies the dielectric response of $BaTiO_3$ therefore remains unclear. Using electric-field-induced molecular dynamics with a machine-learning force field, we show that the permittivity tracks the field-induced angular redistribution of local Ti–O off-centering, not its magnitude. Temperature and biaxial strain modify the local structure through different microscopic routes, yet both dielectric responses collapse onto a common relation with the same orientational descriptor, providing a real-space counterpart to soft-mode behavior.

*Introduction*—The dielectric response of ferroelectric materials varies strongly with temperature and often becomes exceptionally large near ferroelectric phase transitions [1–3]. This behavior is central to ferroelectric physics and directly affects devices such as capacitors and piezoelectrics [4,5]. Yet the real-space atomic degree of freedom underlying this dielectric response remains unclear.

$BaTiO_3$ (BTO), whose ferroelectricity was discovered in the 1940s [6], provides a prototypical system in which to address this question. Upon heating, BTO undergoes successive rhombohedral, orthorhombic, tetragonal, and cubic phase transitions accompanied by pronounced changes in dielectric response [1–3]. Its finite-temperature behavior has long been described within lattice-dynamical theories based on the softening of transverse optical modes [7,8]. Within the framework represented by the Lyddane–Sachs–Teller relation, enhancement of the low-frequency permittivity is associated with a decrease in the frequency of a transverse optical soft mode [9]. At the same time, an order-disorder picture has also been widely discussed, in which local Ti off-centering persists at finite temperature and phase transitions involve changes in the orientational ordering of these local polar distortions [10–12]. BTO therefore exhibits characteristics of both displacive and order-disorder behavior [13,14].

These descriptions characterize either collective lattice dynamics or statistical orientational order, but neither directly specifies how an individual local polar unit responds to an applied field. Equilibrium atomistic simulations have reproduced the temperature-dependent permittivity of BTO from polarization fluctuations [15], but such approaches do not directly resolve the field-induced response of individual local polar units. Recent machine-learning force fields (MLFFs) enable simulations over substantially larger length and time scales [16–19], while MACEField [20] provides instantaneous Born effective charges (BECs), enabling electric-field-induced molecular dynamics (MD) [21–25].

Here we use MACEField to identify the local real-space response underlying the finite-temperature dielectric behavior of BTO. We find that the large variation in permittivity across the tetragonal phase is not accompanied by a comparable change in either the magnitude or the mean polar angle of the local Ti–O off-centering. Instead, it is the field-induced redistribution of its angular population that changes strongly. Biaxial strain modifies the local structure through a different microscopic route, yet the temperature- and strain-dependent dielectric responses collapse onto a common relation with the same orientational response. This identifies the field-induced angular response of the local Ti–O polar unit as a real-space quantity underlying the macroscopic dielectric response.

*Computational approach*—We employed an $r^2$SCAN-based MACEField model selected from LDA-, PBEsol-, and $r^2$SCAN-based models on the basis of their finite-temperature phase-transition behavior [26–28]. Reference configurations were generated from density-functional-theory calculations using VASP [29–32] and its on-the-fly machine-learning framework [33]. BEC tensors were evaluated using PBEsol density-functional perturbation theory [34] for $r^2$SCAN-generated structures and independently validated against $r^2$SCAN finite-field calculations using the perturbation expansion after discretization (PEAD) method [35]. The MACEField model was fine-tuned from MACE-MP-0 large [19]. Model selection and validation and additional analyses are provided in the Supplemental Material [36].

Finite-temperature MD was performed under zero external stress, and the dielectric response was obtained by applying electric fields along the crystallographic $a$ direction, with the forces evaluated from the instantaneous MACEField BEC tensors.

Here, the Cartesian $x$, $y$, and $z$ directions correspond to the crystallographic $a$, $b$, and $c$ directions, respectively. Throughout, T and C denote the tetragonal and cubic phases, respectively. The polarization-switching behavior was independently validated and is consistent with prior experimental and computational studies [37–40]. Full simulation parameters and definitions are given in the End Matter. For each Ti atom, we characterize the local polar structure by the displacement $\boldsymbol{u}$ of Ti relative to the center of its six-neighbor O cage. Its polar angle $\theta$ is measured from the [001] polarization direction. To quantify the field-induced angular response along the crystallographic $a$ direction, we define

$$C_x(E) = \left\langle \frac{u_x}{|\,\boldsymbol{u}\,|} \right\rangle_E = \langle \sin\theta\cos\varphi \rangle_E \tag{1}$$

and the field-induced orientational response

$$R_x = \frac{C_x(+E) - C_x(-E)}{2}. \tag{2}$$

Thus, $R_x$ measures the signed angular redistribution induced by field reversal and can change even when the mean polar angle remains nearly unchanged.

*Finite-temperature dielectric response*—Figure 1(a) shows the relative dielectric constant $\varepsilon_r$ along the crystallographic $a$ direction from the tetragonal into the cubic phase. Because the absolute transition temperatures differ between calculation and experiment, temperature is expressed as $T^T_{\mathrm{norm}}$, such that 0 and 1 correspond to the orthorhombic-to-tetragonal and tetragonal-to-cubic transitions, respectively. The calculated $\varepsilon_r$ decreases markedly with increasing temperature over most of the tetragonal region and exhibits a sharp enhancement on approaching the Curie temperature, while the calculated values at $T^T_{\mathrm{norm}} = 1.16$ and 1.42 also closely reproduce the experimental dielectric response in the cubic phase [1–3].

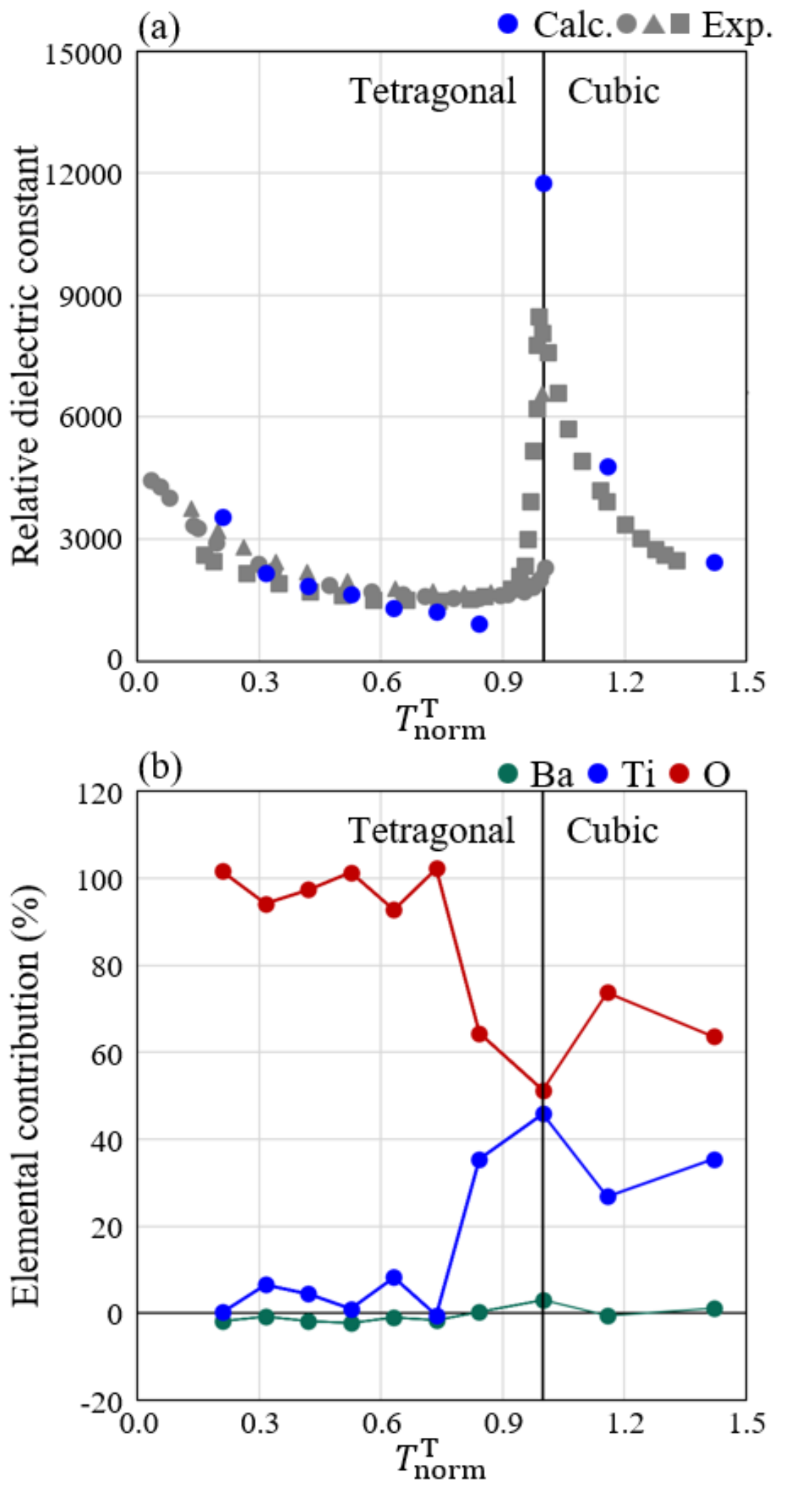

FIG. 1. **Finite-temperature dielectric response of BTO across the tetragonal and cubic phases and its element-resolved contributions.** (a) Relative dielectric constant $\varepsilon_r$ as a function of the normalized temperature $T^{\mathrm{T}}_{\mathrm{norm}}$. The calculated results are shown by blue circles, while experimental data from Refs. [1–3] are shown by gray circles, squares, and triangles, respectively. $T^{\mathrm{T}}_{\mathrm{norm}}$=0 and 1 correspond to the O–T and T–C phase transitions, respectively, and $T^T_{\mathrm{norm}} > 1$ denotes temperatures above the T–C transition. (b) Element-resolved contributions of Ba, Ti, and O to the polarization-derived dielectric response. O dominates the response on the lower-temperature side of the tetragonal phase, whereas the relative contribution of Ti increases toward the Curie temperature and remains substantial above the T–C transition. The contribution of Ba remains small over the temperature range examined. In both panels, the black vertical solid line marks the T–C phase transition at $T^T_{\mathrm{norm}} = 1$.

*Contact author: teru@iis.u-tokyo.ac.jp

The agreement with experiment allows the trajectories to be used to identify the corresponding atomic response. We first decompose the polarization-derived response into contributions from Ba, Ti, and O, as shown in Fig. 1(b). Although Ti off-centering is commonly used to describe the local polar distortion of BTO [13,14], the O sublattice provides the dominant contribution on the lower-temperature side of the tetragonal phase. The Ti contribution is smaller at lower temperature but increases markedly toward the Curie temperature and remains large above the T–C transition, whereas the Ba contribution remains small. A further decomposition of the O response according to Ti–O bond orientation shows that the O site bonded parallel to the applied field dominates the O contribution, consistent with the well-established anisotropy of the O BEC in BTO (Fig. S7 [36] ) [41]. The dielectric response is therefore governed predominantly by relative motion within the Ti–O sublattice.

This result motivates the Ti displacement relative to its surrounding O cage as a natural local coordinate.

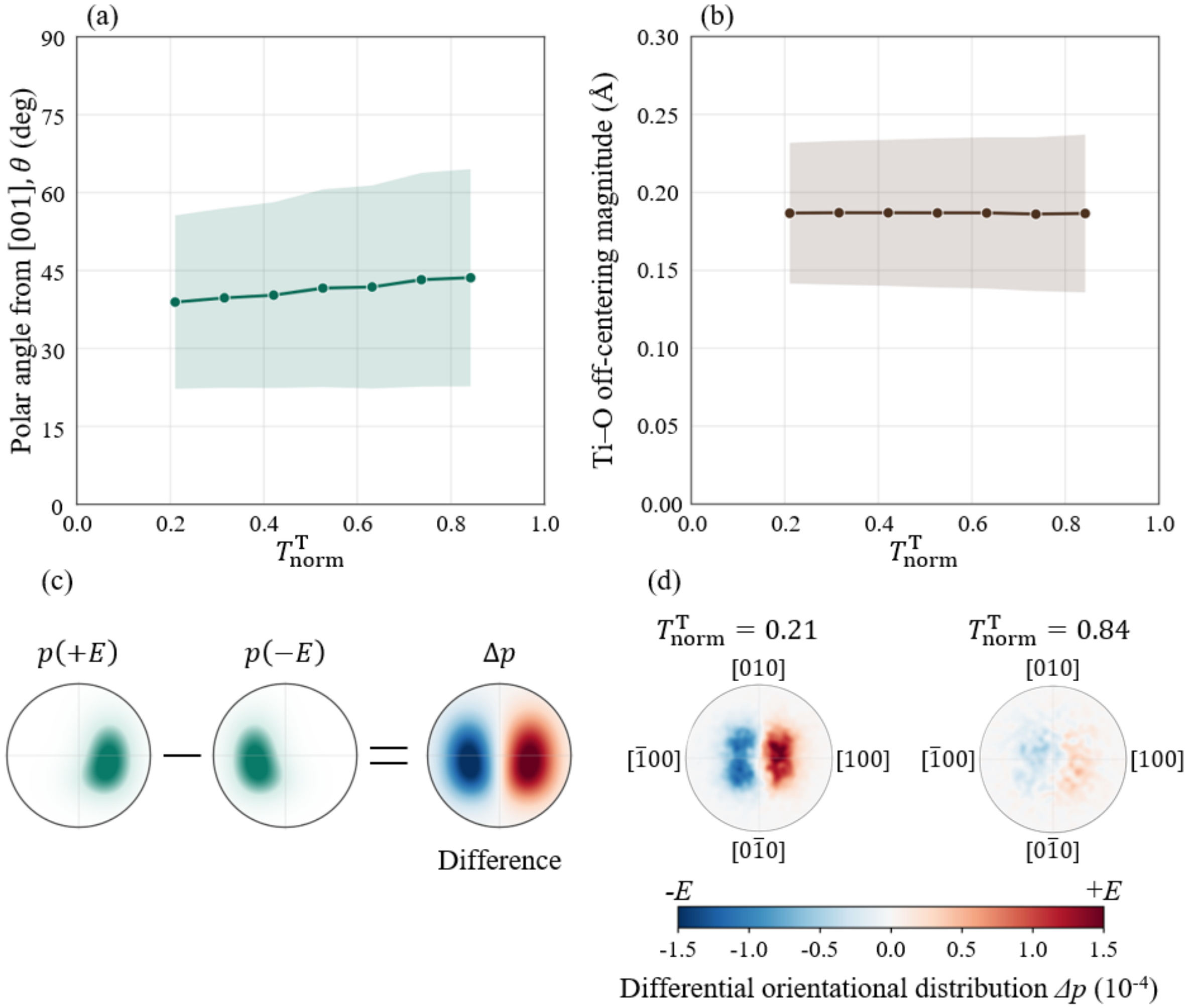

FIG. 2. **Temperature dependence of the local Ti–O polar structure and its electric-field-induced orientational response in tetragonal BTO.** (a) Temperature dependence of the mean polar angle of the Ti–O off-centering with respect to the [001] direction. (b) Temperature dependence of the Ti–O off-centering magnitude. In (a) and (b), the quantities were evaluated from the local structures sampled throughout the entire hysteresis loop; symbols indicate mean values and shaded regions represent the corresponding standard deviations. Panels (c) and (d) use a pole-figure representation of the Ti–O off-centering projected along [001], in which the radial coordinate is the polar angle $\theta$ from [001] and the azimuthal coordinate is the angle $\varphi$ about [001]; the crystallographic a direction lies horizontal, and the electric field is applied parallel to it with opposite polarities. (c) Schematic illustration of the differential orientational analysis. The differential distribution is defined as $\Delta p = p(+E) - p(-E)$, where $p(+E)$ and $p(-E)$ denote the orientational probability distributions under the two field polarities. (d) Differential pole figures at $T^{\mathrm{T}}_{\mathrm{norm}} = 0.21$ and 0.84. Red and blue indicate positive and negative values of $\Delta p$, respectively.

*Contact author: teru@iis.u-tokyo.ac.jp

Unlike the absolute Ti displacement, this coordinate incorporates the motion of both sublattices that dominate the dielectric response. To isolate the microscopic origin of the temperature-dependent response from the change in average crystal symmetry at the T–C transition, we first focus on the tetragonal phase. We next ask which property of this local Ti–O polar unit—its amplitude, its mean orientation, or its response to the applied field—tracks the macroscopic permittivity.

*Local angular response*—Figures 2(a) and 2(b) show the mean polar angle from [001] and the magnitude of the Ti–O off-centering across the tetragonal phase. Despite the pronounced decrease in $\varepsilon_r$ , neither quantity shows a corresponding systematic temperature dependence. The local polar distortion therefore neither appreciably shrinks nor develops a systematic change in its mean inclination as the dielectric response decreases.

The absence of a change in the mean polar angle, however, does not imply that the angular distribution is insensitive to the field. Figure 2(c) illustrates the definition of the differential orientational distribution obtained from trajectories under fields of opposite sign, and Fig. 2(d) compares representative distributions at $T_{\mathrm{norm}}^{\mathrm{T}}$ = 0.21 and 0.84. At the lower temperature, the positive and negative differential lobes appear on opposite sides of the distribution at similar radial positions. The applied field therefore primarily redistributes the local Ti–O orientations azimuthally about [001], rather than changing their inclination from the polarization axis. This redistribution becomes markedly weaker at higher temperature.

Correspondingly, Rx decreases from 0.0676 at 260 K to 0.0178 at 320 K — a factor of 3.8 — whereas the mean polar angle changes only from 38.9° to 43.6° and the off-centering magnitude only from 0.1868 to 0.1865 Å.

*Strain control*—If this orientational response provides the relevant local descriptor of the permittivity, it should also track the dielectric response when that response is modified by a perturbation other than temperature. We therefore applied equal biaxial strain along the crystallographic $a$ and $b$ axes at a fixed temperature of 290 K, while the dielectric response was probed by applying the electric field along the crystallographic $a$ direction.

As shown in Fig. 3, $\varepsilon_r$ increases from 386 under $-0.3\%$ (compressive) strain to 2559 under $+0.3\%$ (tensile) strain. Thus, even at fixed temperature, the dielectric response can be varied by a factor of 6.6 through a strain change of only 0.6%.

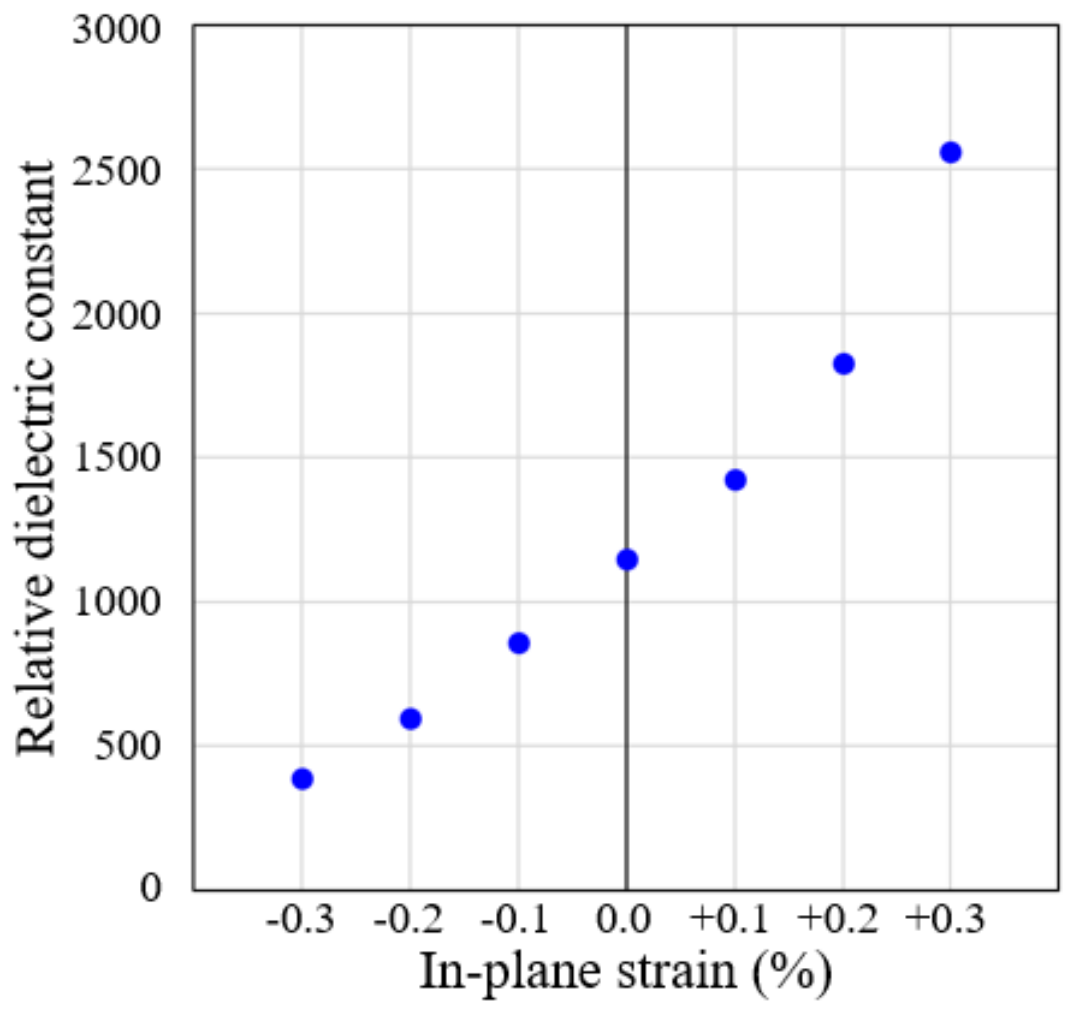

FIG. 3. **Strain modulation of the dielectric response in tetragonal BTO.** Relative dielectric constant $\varepsilon_r$ as a function of equal biaxial strain along the crystallographic $a$- and $b$-axes at a fixed temperature of 290 K, with the electric field applied along the crystallographic $a$ direction. The dielectric response increases systematically from compressive to tensile strain, rising from 386 at $-0.3\%$ strain to 2559 at $+0.3\%$ strain.

The corresponding local response differs qualitatively from the temperature-dependent case. Figures 4(a) and 4(b) show that tensile strain increases the mean polar angle of the Ti–O distortion from 34.9° to 53.7°, whereas its magnitude varies by less than 5%. Strain therefore modifies the mean inclination of the local polar units in addition to changing their field response. The differential orientational distributions in Fig. 4(c) reveal a concomitant increase in the field-induced angular redistribution. Under $-0.3\%$ (compressive) strain the redistribution is weak, whereas under $+0.3\%$ (tensile) strain it becomes pronounced. Consistently, $R_x$ increases from 0.0091 to 0.0588. Temperature and strain therefore act through distinct microscopic routes: temperature primarily modifies the field-induced angular redistribution while leaving the mean polar angle nearly unchanged, whereas strain modifies both the mean orientation and its field response.

*Contact author: teru@iis.u-tokyo.ac.jp

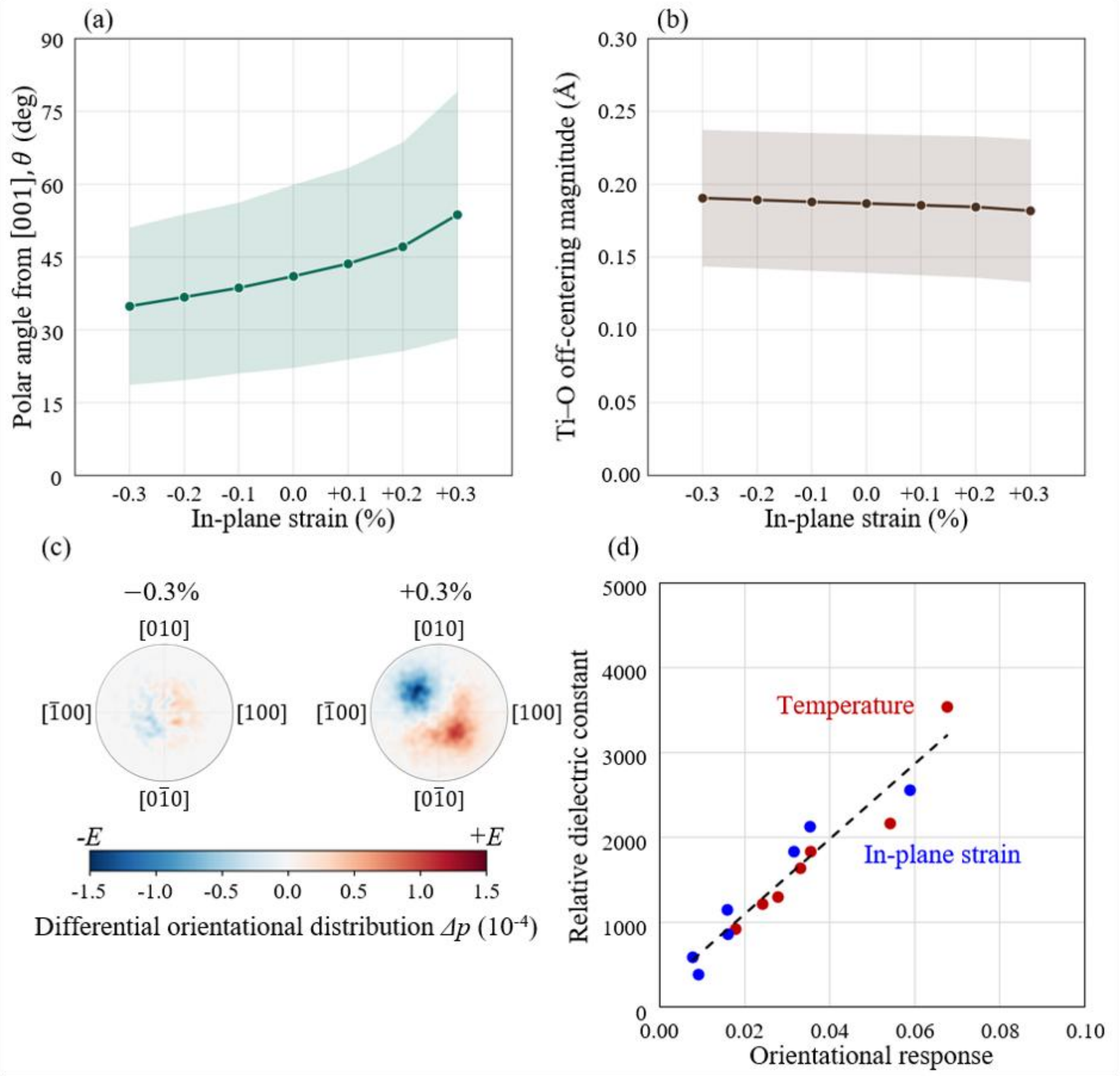

FIG. 4. **Strain dependence of the local Ti–O polar structure and common relationship between the dielectric and orientational responses.** Equal biaxial strain is applied along the crystallographic $a$ and $b$ axes, while the electric field is applied along the crystallographic $a$ direction. (a) Strain dependence of the mean polar angle of the Ti–O off-centering with respect to [001]. (b) Strain dependence of the Ti–O off-centering magnitude. The quantities in (a) and (b) were evaluated from the local structures sampled throughout the entire hysteresis loop. Symbols indicate mean values and shaded regions represent the corresponding standard deviations. (c) Differential pole figures of the Ti–O off-centering viewed along the [001] direction at biaxial in-plane strains of $-0.3\%$ and $+0.3\%$. The difference between the orientational probability distributions under $+E$ and $-E$ is shown. Red and blue indicate an increase and decrease, respectively, in occupation probability under $+E$ relative to $-E$. (d) Relative dielectric constant $\varepsilon_r$ plotted against the orientational response $R_x$ for the temperature-dependent and in-plane-strain-dependent datasets. Temperature- and strain-dependent data are shown by red and blue symbols, respectively. The dashed line represents a linear regression to the combined dataset, $\varepsilon_r = 4.42 \times 10^4 R_x + 210$, with $R^2 = 0.925$.

*Angular response and soft-mode behavior*—Despite these distinct microscopic routes, the dielectric responses produced by temperature and strain are described by the same local quantity. Figure 4(d) compares $\varepsilon_r$ directly with $R_x$. The temperature- and strain-dependent datasets collapse onto a common linear relation, with a combined regression $\varepsilon_r = 4.42 \times 10^4 R_x + 210$ and $R^2 = 0.925$. This collapse is not tautological: $R_x$ is constructed from unit off-centering vectors and therefore carries no information about the magnitude of the local distortion or about the BEC weighting that determines P. Indeed, the dominant contribution to $\varepsilon_r$ arises from the O sublattice (Fig. 1(b)), whereas $R_x$ is defined from the orientation of the Ti–O unit, and neither the mean polar angle nor the off-centering magnitude produces such a collapse. Additional temperature-dependent calculations under the same in-plane lattice constraint used for the strain calculations retain both the characteristic temperature dependence and a similar $\varepsilon_r - R_x$ relationship [36], showing that the correlation is not an artifact of the different lattice constraints.

*Contact author: teru@iis.u-tokyo.ac.jp

The finite-temperature dielectric response of BTO has been quantitatively described for decades in terms of soft transverse optical modes and the Lyddane–Sachs–Teller relation [9,42]. A mode frequency, however, is a spectral description of collective dynamics and does not by itself specify how that softening is manifested in the real-space response of individual polar units. This distinction is particularly relevant in tetragonal BTO, where the low-frequency $E(\mathrm{TO})$ soft mode associated with the $a$-axis response is strongly overdamped [43].

Our results provide such a real-space description. As the permittivity decreases with temperature, the local Ti–O distortion neither appreciably shrinks nor systematically tilts; rather, it becomes less responsive to field-induced angular redistribution. In this sense, the relevant local softness is angular. This picture is complementary to, rather than an alternative to, the conventional soft-mode description. It also gives a field-response interpretation to the order-disorder character of BTO: the local polar distortions persist while their field-induced angular response changes. At 350 and 375 K, the mean local Ti–O off-centering magnitude remains approximately 0.175 Å, whereas the corresponding vector-averaged magnitudes are only 6.7% and 4.1% of the mean local magnitudes, respectively, indicating that the local distortion persists while differently oriented off-centering vectors largely cancel upon averaging [36]. Importantly, $\varepsilon_r$ continues to track $R_x$ above the transition [36], showing that the same local orientational response remains relevant to the dielectric behavior in both the tetragonal and cubic phases.

Within the tetragonal phase, the common $\varepsilon_r - R_x$ relation further shows that this field-induced angular response is not determined uniquely by temperature. Strain accesses the same relation from a different local structural distribution. The field-induced angular response of local polar units is therefore a tunable microscopic property and provides a possible design coordinate for controlling dielectric response independently of the magnitude of the local polar distortion.

*Conclusion*— The dielectric response of tetragonal BTO is governed by the field-induced angular response of its local Ti–O polar structure. Temperature and biaxial strain modify the underlying local structure through different microscopic routes, yet both dielectric responses track the same orientational response $R_x$. Neither the magnitude of the local polar distortion nor its mean tilt from [001] shows the same correspondence with the permittivity. Within the tetragonal phase the two datasets follow a single linear relation; across the T–C transition Rx continues to track the dielectric response, although the slope and intercept change.

The resulting picture separates the existence of a local polar distortion from its orientability. In BTO, the enhanced dielectric response is associated not with an increase in the magnitude of the local polar distortion but with a stronger field-induced orientational response. Orientability is therefore a design coordinate: it can be tuned by strain at fixed temperature, and it can be evaluated for any polar material directly from trajectories under reversed electric fields.

*Acknowledgments*—This study was supported by the Ministry of Education, Culture, Sports, Science and Technology (MEXT, 26K22608, 26K01205, 25K24645). P.-Y.C. acknowledges support from JST SPRING (Grant No. JPMJSP2108). The computations were performed using resources provided by Genkai, the Research Institute for Information Technology, Kyushu University.

R.S. and P.-Y.C. conceived the idea and designed the research. R.S. performed the simulations, analyzed the data, and prepared the figures. R.S. wrote the original draft of the manuscript. T.M. supervised the project and acquired funding. All authors discussed the results and reviewed and edited the manuscript.

The authors declare no competing financial interest.

*Data availability*—The trained machine-learning force field models, Born effective charge models, and simulation trajectories generated in this study are available from the corresponding author upon reasonable request.

[1] S. H. Wemple, M. DiDomenico, and I. Camlibel, Dielectric and optical properties of melt-grown BaTiO3, J. Phys. Chem. Solids **29**, 1797 (1968).

*Contact author: teru@iis.u-tokyo.ac.jp

[2] C. J. Johnson, Some dielectric and electro-optic properties of $BaTiO_3$ single crystals, Appl. Phys. Lett. **7**, 221 (1965).
[3] I. Camlibel, M. DiDomenico, and S. H. Wemple, Dielectric properties of single-domain melt-grown BaTiO3, J. Phys. Chem. Solids **31**, 1417 (1970).
[4] K. Hong, T. H. Lee, J. M. Suh, S.-H. Yoon, and H. W. Jang, Perspectives and challenges in multilayer ceramic capacitors for next generation electronics, J. Mater. Chem. C **7**, 9782 (2019).
[5] N. Iqbal, A. Dixit, P. S. Dobal, R. S. Katiyar, and A. S. Bhalla, Doping effects in barium titanate: a historical perspective and review, J. Mater. Sci.: Mater. Electron. **36**, 1008 (2025).
[6] B. Wul, Barium titanate: a new ferro-electric, Nature **157**, 808 (1946).
[7] J. Petzelt, Soft mode behavior in cubic and tetragonal $BaTiO_3$ crystals and ceramics: Review on the results of dielectric spectroscopy, Ferroelectrics **375**, 156 (2008).
[8] S. Kamba, Soft-mode spectroscopy of ferroelectrics and multiferroics: A review, APL Mater. **9**, 020704 (2021).
[9] R. H. Lyddane, R. G. Sachs, and E. Teller, On the polar vibrations of Alkali Halides, Phys. Rev. **59**, 673 (1941).
[10] X.-G. Zhao, O. I. Malyi, S. J. L. Billinge, and A. Zunger, Intrinsic local symmetry breaking in nominally cubic paraelectric $BaTiO_3$, Phys. Rev. B **105**, 224108 (2022).
[11] S. Sanna, C. Thierfelder, S. Wippermann, T. P. Sinha, and W. G. Schmidt, Barium titanate ground- and excited-state properties from first-principles calculations, Phys. Rev. B **83**, 054112 (2011).
[12] R. Comes, M. Lambert, and A. Guinier, The chain structure of $BaTiO_3$ and $KNbO_3$, Solid State Commun. **6**, 715 (1968).
[13] B. Zalar, V. V. Laguta, and R. Blinc, NMR evidence for the coexistence of order-disorder and displacive components in barium titanate, Phys. Rev. Lett. **90**, 037601 (2003).
[14] M. S. Senn, D. A. Keen, T. C. A. Lucas, J. A. Hriljac, and A. L. Goodwin, Emergence of long-range order in $BaTiO_3$ from local symmetry-breaking distortions, Phys. Rev. Lett. **116**, 207602 (2016).
[15] L. Gigli, M. Veit, M. Kotiuga, G. Pizzi, N. Marzari, and M. Ceriotti, Thermodynamics and dielectric response of $BaTiO_3$ by data-driven modeling, npj Comput. Mater. **8**, 209 (2022).
*Contact author: teru@iis.u-tokyo.ac.jp
[16] P.-Y. Chen, K. Shibata, and T. Mizoguchi, High precision machine learning force field development for $BaTiO_3$ phase transitions, amorphous, and liquid structures, APL Mach. Learn. **3**, 036115 (2025).
[17] J. Zhang, H. Zhang, H. Zheng, B. Xu, J. Wang, and X. Guo, On-the-fly machine learning-assisted high accuracy second-principles model for $BaTiO_3$, npj Comput. Mater. **11**, 299 (2025).
[18] I. Batatia, D. P. Kovács, G. Simm, C. Ortner, and G. Csányi, MACE: Higher order equivariant message passing neural networks for fast and accurate force fields, in *Advances in Neural Information Processing Systems 35* (Neural Information Processing Systems Foundation, Inc., San Diego, CA, 2022), pp. 11423–11436.
[19] I. Batatia et al., A foundation model for atomistic materials chemistry, J. Chem. Phys. **163**, 184110 (2025).
[20] B. A. A. Martin, A. M. Ganose, V. Kapil, T. Li, and K. T. Butler, General learning of the electric response of inorganic materials, PRX Intelligence **1**, 013006 (2026).
[21] R. Sahashi, P.-Y. Chen, and T. Mizoguchi, LDA-based machine learning force field for accurate electric-field-driven ferroelectric response in $BaTiO_3$, J. Ceram. Soc. Jpn. **134**, 431 (2026).
[22] P.-Y. Chen and T. Mizoguchi, Effect of uniaxial compressive stress on polarization switching and domain wall formation in tetragonal phase $BaTiO^3$ via machine learning potential, Mater. Des. **265**, 115851 (2026).
[23] P.-Y. Chen and T. Mizoguchi, Electric field-induced phase transitions and hysteresis in ferroelectric $HfO_2$ captured with machine learning potential, Mater. Today Electron. **17**, 100228 (2026).
[24] P.-Y. Chen and T. Mizoguchi, Long-range interaction effects on the phase transition, mechanical effect, and electric field response of $BaTiO_3$ by machine learning potentials, APL Mach. Learn. **4**, 036101 (2026).
[25] P.-Y. Chen and T. Mizoguchi, Transition from homogeneous to domain-wall-mediated polarization switching in $BaTiO_3$: A machine-learning molecular dynamics study, Phys. Rev. B (to be published).
[26] J. P. Perdew and A. Zunger, Self-interaction correction to density-functional approximations for many-electron systems, Phys. Rev. B **23**, 5048 (1981).

[27] J. P. Perdew, A. Ruzsinszky, G. I. Csonka, O. A. Vydrov, G. E. Scuseria, L. A. Constantin, X. Zhou, and K. Burke, Restoring the density-gradient expansion for exchange in solids and surfaces, Phys. Rev. Lett. **100**, 136406 (2008).
[28] J. W. Furness, A. D. Kaplan, J. Ning, J. P. Perdew, and J. Sun, Accurate and numerically efficient r$^2$SCAN meta-generalized gradient approximation, J. Phys. Chem. Lett. **11**, 8208 (2020).
[29] G. Kresse and J. Hafner, *Ab initio* molecular dynamics for liquid metals, Phys. Rev. B **47**, 558 (1993).
[30] G. Kresse and J. Furthmüller, Efficiency of ab-initio total energy calculations for metals and semiconductors using a plane-wave basis set, Comput. Mater. Sci. **6**, 15 (1996).
[31] G. Kresse and J. Furthmüller, Efficient iterative schemes for *ab initio* total-energy calculations using a plane-wave basis set, Phys. Rev. B **54**, 11169 (1996).
[32] G. Kresse and D. Joubert, From ultrasoft pseudopotentials to the projector augmented-wave method, Phys. Rev. B **59**, 1758 (1999).
[33] R. Jinnouchi, F. Karsai, and G. Kresse, On-the-fly machine learning force field generation: Application to melting points, Phys. Rev. B **100**, 014105 (2019).
[34] X. Gonze and C. Lee, Dynamical matrices, Born effective charges, dielectric permittivity tensors, and interatomic force constants from density-functional perturbation theory, Phys. Rev. B **55**, 10355 (1997).
[35] R. W. Nunes and X. Gonze, Berry-phase treatment of the homogeneous electric field perturbation in insulators, Phys. Rev. B **63**, 155107 (2001).
[36] See Supplemental Material at [URL will be inserted by publisher] for model selection and validation, phase-transition identification, polarization-switching validation, constrained-lattice calculations, bond-orientation-resolved oxygen contributions, and the extension of the local-response analysis across the T–C transition.
[37] W. J. Merz, The electric and optical behavior of $BaTiO_3$ single-domain crystals, Phys. Rev. **76**, 1221 (1949).
[38] B. Jiang, Y. Bai, W. Chu, Y. Su, and L. Qiao, Direct observation of two 90° steps of 180° domain switching in $BaTiO_3$ single crystal under an antiparallel electric field, Appl. Phys. Lett. **93**, 152905 (2008).
[39] H. Azuma, T. Ogawa, S. Ogata, R. Kobayashi, M. Uranagase, T. Tsuzuki, and F. Wendler, Unique temperature-dependence of polarization switching paths in ferroelectric $BaTiO_3$: A molecular dynamics simulation study, Acta Mater. **296**, 121216 (2025).
[40] Y. Li, J. Wang, and F. Li, Intrinsic polarization switching in $BaTiO_3$ crystal under uniaxial electromechanical loading, Phys. Rev. B **94**, 184108 (2016).
[41] Ph. Ghosez, J.-P. Michenaud, and X. Gonze, Dynamical atomic charges: The case of $ABO_3$ compounds, Phys. Rev. B **58**, 6224 (1998).
[42] G. Shirane, J. D. Axe, J. Harada, and A. Linz, Inelastic neutron scattering from single-domain $BaTiO_3$, Phys. Rev. B **2**, 3651 (1970).
[43] G. Burns and F. H. Dacol, Lattice modes in ferroelectric perovskites. III. Soft modes in $BaTiO_3$, Phys. Rev. B **18**, 5750 (1978).

End Matter

*Computational methods*—The r$^2$SCAN reference dataset was constructed from density-functional-theory calculations performed using VASP. Forty-atom supercells representing the rhombohedral, orthorhombic, tetragonal, and cubic phases were employed, and finite-temperature configurations were sampled using the on-the-fly machine-learning framework implemented in VASP. Reference energies and atomic forces were evaluated using the r$^2$SCAN meta-GGA functional. Because density-functional perturbation theory for BECs is not currently available for meta-GGA functionals in VASP, the BEC tensors used to train the r$^2$SCAN-based MACEField model were evaluated using PBEsol density-functional perturbation theory (DFPT) for structures generated at the r$^2$SCAN level. This treatment was independently assessed using r$^2$SCAN finite-field calculations based on the perturbation expansion after discretization (PEAD) method. The MACEField model was fine-tuned from MACE-MP-0 large.

Finite-temperature heating simulations were initialized from the rhombohedral phase after full relaxation of the atomic positions and lattice parameters. Variable-cell MD was performed under zero external stress. The thermostat and barostat time constants were 25 and 75 fs, respectively, and a time step of 1 fs was used. The system was heated linearly from 1 to 500 K over 200 ps, and atomic coordinates and lattice parameters were recorded every 1 ps.

For Ti atom $i$, the local off-centering vector was defined relative to the center of its six nearest-neighbor O atoms as

$$\boldsymbol{u}_i = \boldsymbol{r}_{\mathrm{Ti},i} - \frac{1}{6}\sum_{j=1}^{6} \boldsymbol{r}_{\mathrm{O},ij} \,. \qquad (A1)$$

*Contact author: teru@iis.u-tokyo.ac.jp

Its polar angle was defined with respect to [001] as $\theta = \cos^{-1}(u_z/|\boldsymbol{u}|)$. Phase-transition temperatures were identified using both the reduced lattice parameters and changes in the mean orientation of the Ti–O off-centering. A Gaussian filter was applied to the temperature-dependent orientation angles, and their temperature derivatives were evaluated to identify candidate transitions. Final transition temperatures were assigned only when consistent anomalies were observed in both the lattice parameters and multiple orientational components.

Electric-field-induced molecular dynamics simulations were performed on an 8 × 8 × 8 $BaTiO_3$ supercell containing 2560 atoms in the NPT ensemble, using a Nosé–Hoover thermostat and a Parrinello–Rahman barostat in the Melchionna formulation. The thermostat time constant was 25 fs, and the barostat time parameter was set to 75 fs. A time step of 1 fs was used. Initial configurations were prepared by zero-field NPT equilibration for 50 ps at each target temperature. A triangular electric field was then applied along the crystallographic a direction, following the sequence 0 → −10 → +10 → 0 kV $cm^{-1}$ over 80 ps, corresponding to a sweep rate of $5.0 \times 10^{-4}$ kV $cm^{-1}$ $fs^{-1}$. The external forces were evaluated using configuration-dependent BEC tensors predicted by MACEField and updated every 10 MD steps. Polarization was recorded every 100 steps. The relative permittivity along the field direction was estimated from a linear fit to the polarization data over the entire hysteresis cycle.

For the strain-dependent calculations, zero strain refers to the zero-stress equilibrium in-plane lattice constant obtained at 290 K, and the a and b axes were then constrained to ±0.1%, ±0.2%, and ±0.3% of that value.The external force acting on atom $i$ was evaluated as

$$F_{i,\alpha}^{\mathrm{ext}} = |e| \sum_{\beta} E_\beta Z_{i,\alpha\beta}^{*}, \qquad (A2)$$

where $E_\beta$ is the applied electric-field component and $\boldsymbol{Z}_{i,\alpha\beta}^{*}$ is the instantaneous BEC tensor.

The macroscopic polarization was evaluated relative to a nonpolar cubic reference structure using

$$\boldsymbol{P} = \frac{|e|}{V} \sum_{i} \boldsymbol{Z}_i^{*} \Delta \boldsymbol{u}_i, \qquad (A3)$$

where $V$ is the instantaneous cell volume and $\Delta\boldsymbol{u}_i$ is the displacement of atom $i$ from its corresponding reference position. The relative dielectric constant along the crystallographic $a$ direction was obtained from the low-field polarization slope,

$$\varepsilon_{r,a} = 1 + \frac{1}{\varepsilon_0} \frac{dP_a}{dE_a}. \qquad (A4)$$

To resolve the response by atomic species, the polarization sum in Eq. (A3) was restricted to species $s$. The corresponding polarization-response contribution was

$$\varepsilon_{a,s} = \frac{1}{\varepsilon_0} \frac{dP_{a,s}}{dE_a}, \qquad (A5)$$

and its fractional contribution was evaluated as

$$C_s = \frac{\varepsilon_{a,s}}{\sum_s \varepsilon_{a,s}} \times 100. \qquad (A6)$$

The vacuum contribution of unity in Eq. (A4) was not assigned to any atomic species.

For the orientational analysis, the [001] direction was used as the polar axis. The polar angle $\theta$ measures the inclination of the Ti–O off-centering vector from [001], while the azimuthal angle $\varphi$ describes rotation about [001]. With $E \parallel [100]$, the orientational response was evaluated using Eqs. (1) and (2). Because $C_x$ depends on both $\theta$ and $\phi$, $R_x$ can change through redistribution in the azimuthal coordinate even when the mean polar angle remains nearly constant.

For the strain-dependent calculations, the $a$ and $b$ lattice axes were constrained to the same prescribed biaxial strain, while the dielectric and local structural responses were evaluated at a fixed temperature of 290 K. Additional temperature-dependent simulations employing the same in-plane constraint were performed to verify that neither the characteristic temperature dependence nor the $\varepsilon_r - R_x$ relationship originated from the different lattice constraints used in the temperature- and strain-dependent calculations.

*Contact author: teru@iis.u-tokyo.ac.jp

**Supplemental Material for: Angular compliance, not amplitude, governs the dielectric response of $BaTiO_3$**

Ryotaro Sahashi[1], Po-Yen Chen[1], and Teruyasu Mizoguchi[1,2,*]

[1]Department of Materials Engineering, The University of Tokyo, Tokyo, Japan
[2]Institute of Industrial Science, The University of Tokyo, Tokyo, Japan
[*]Contact author: teru@iis.u-tokyo.ac.jp

This Supplemental Material provides details of the selection and validation of the $r^2$SCAN-based MACEField model, including the exchange-correlation-functional dependence of the finite-temperature phase-transition behavior, validation of Born effective charges (BECs) and phonon dispersions, the procedure used to identify the phase-transition temperatures, validation of ferroelectric polarization switching, additional dielectric-response calculations under constrained in-plane lattice conditions, a bond-orientation-resolved analysis of the oxygen contribution, and an extension of the local-response analysis across the T–C transition.

## SI. Selection of the $r^2$SCAN-Based MACEField Model

Three exchange-correlation-functional-dependent models were considered to select the model used for the subsequent finite-temperature and electric-field-coupled molecular dynamics (MD) simulations. For the LDA-based model, the reference dataset reported in Ref. [1] was employed and the MACEField model was constructed in the present study. For PBEsol, the previously reported trained model of Ref. [2] was used directly without additional fine-tuning. In contrast, the $r^2$SCAN reference dataset and the corresponding $r^2$SCAN-based MACEField model were constructed in the present study.

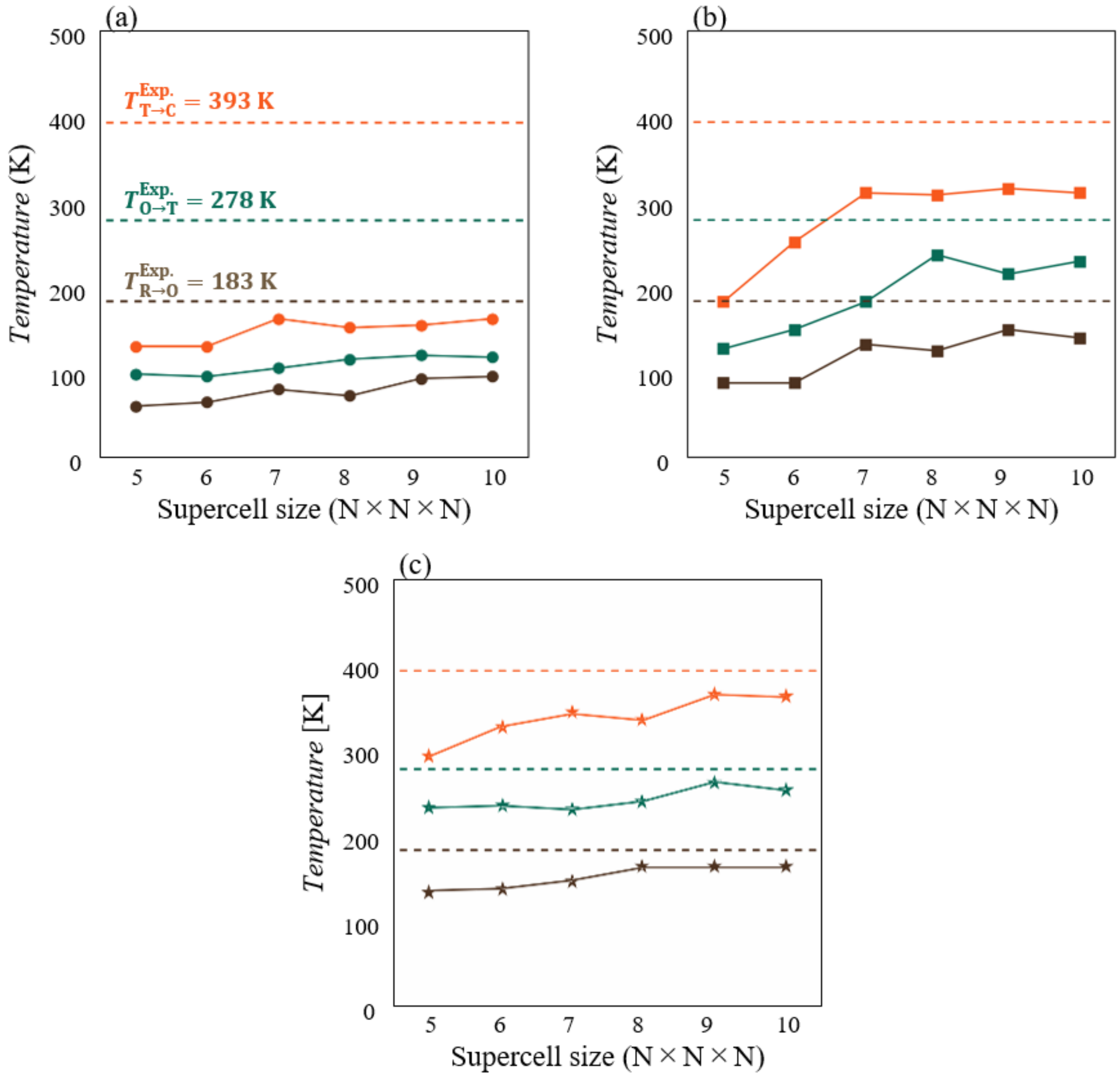

FIG. S1. Exchange-correlation-functional and supercell-size dependence of the finite-temperature phase transitions of BTO. Calculated R–O, O–T, and T–C transition temperatures obtained using the (a) LDA-, (b) PBEsol-, and (c) $r^2$SCAN-based MACEField models as a function of supercell size. Solid lines represent the calculated transition temperatures, and horizontal dashed lines indicate the corresponding experimental values reported in Ref. [3]. The R–O, O–T, and T–C transitions are shown in brown, green, and orange, respectively. The LDA- and PBEsol-based models systematically underestimate the transition temperatures, whereas the $r^2$SCAN-based model provides substantially improved agreement with experiment.

The rhombohedral, orthorhombic, tetragonal, and cubic phases are hereafter denoted R, O, T, and C, respectively. Because the central purpose of the present work is to examine finite-temperature dielectric properties, the models were compared primarily in terms of their ability to reproduce the sequence and temperatures of the R–O, O–T, and T–C phase transitions of $BaTiO_3$ (BTO). Figure S1 shows the calculated transition temperatures as a function of supercell size for the LDA-, PBEsol-, and $r^2$SCAN-based models. The corresponding experimental values reported in Ref. [3] are included for comparison.

Both the LDA- and PBEsol-based models systematically underestimate the experimental transition temperatures. In contrast, the $r^2$SCAN-based model yields substantially improved transition temperatures and reproduces the experimentally observed R–O–T–C sequence over the supercell sizes examined. The transition temperatures averaged over the examined supercell sizes and their deviations from experiment are summarized in Table SI. The $r^2$SCAN-based model gives the smallest overall MAE, 41.1 K, and was therefore selected for the electric-field-coupled simulations and local-response analyses presented in the main text. To our knowledge, simultaneous quantitative reproduction of all three R–O, O–T, and T–C transition temperatures together with the temperature-dependent dielectric response has not been reported previously (see, e.g., Ref. [4]).

*Contact author: teru@iis.u-tokyo.ac.jp

TABLE SI. Comparison of the finite-temperature phase-transition temperatures predicted by the LDA-, PBEsol-, and r$^2$SCAN-based MACEField models. Mean R–O, O–T, and T–C transition temperatures and their standard deviations are evaluated over the supercell sizes examined in Fig. S1 (5 × 5 × 5 to 10 × 10 × 10). The overall mean absolute error (MAE) is calculated relative to the corresponding experimental transition temperatures and is used as the primary criterion for model selection.

| Model | LDA | PBEsol | r$^2$SCAN |
|---|---|---|---|
| $T_{\mathrm{R}\rightarrow\mathrm{O}}$ (K) | 77.5 ± 14.3 | 120.4 ± 26.8 | 151.3 ± 13.0 |
| $T_{\mathrm{O}\rightarrow\mathrm{T}}$ (K) | 108.3 ± 10.7 | 190.4 ± 44.8 | 242.1 ± 12.8 |
| $T_{\mathrm{T}\rightarrow\mathrm{C}}$ (K) | 148.8 ± 15.1 | 279.6 ± 53.0 | 337.5 ± 26.6 |
| MAE (K) | 173.1 | 87.9 | 41.1 |

## SII. Validation of the r$^2$SCAN-Based MACEField Model

The r$^2$SCAN-based model selected from the phase-transition analysis was further validated against first-principles reference data for energies, atomic forces, BECs, and lattice dynamics.

Because density-functional perturbation theory (DFPT) calculations of BECs are not currently implemented for meta-GGA functionals in VASP, the BEC tensors used in the r$^2$SCAN-based training dataset were calculated using PBEsol-DFPT for structures generated at the r$^2$SCAN level. To evaluate the validity of this treatment, BECs obtained using PBEsol-DFPT were compared with values independently calculated using the finite-electric-field perturbation-expansion-after-discretization (PEAD) method with the r$^2$SCAN functional.

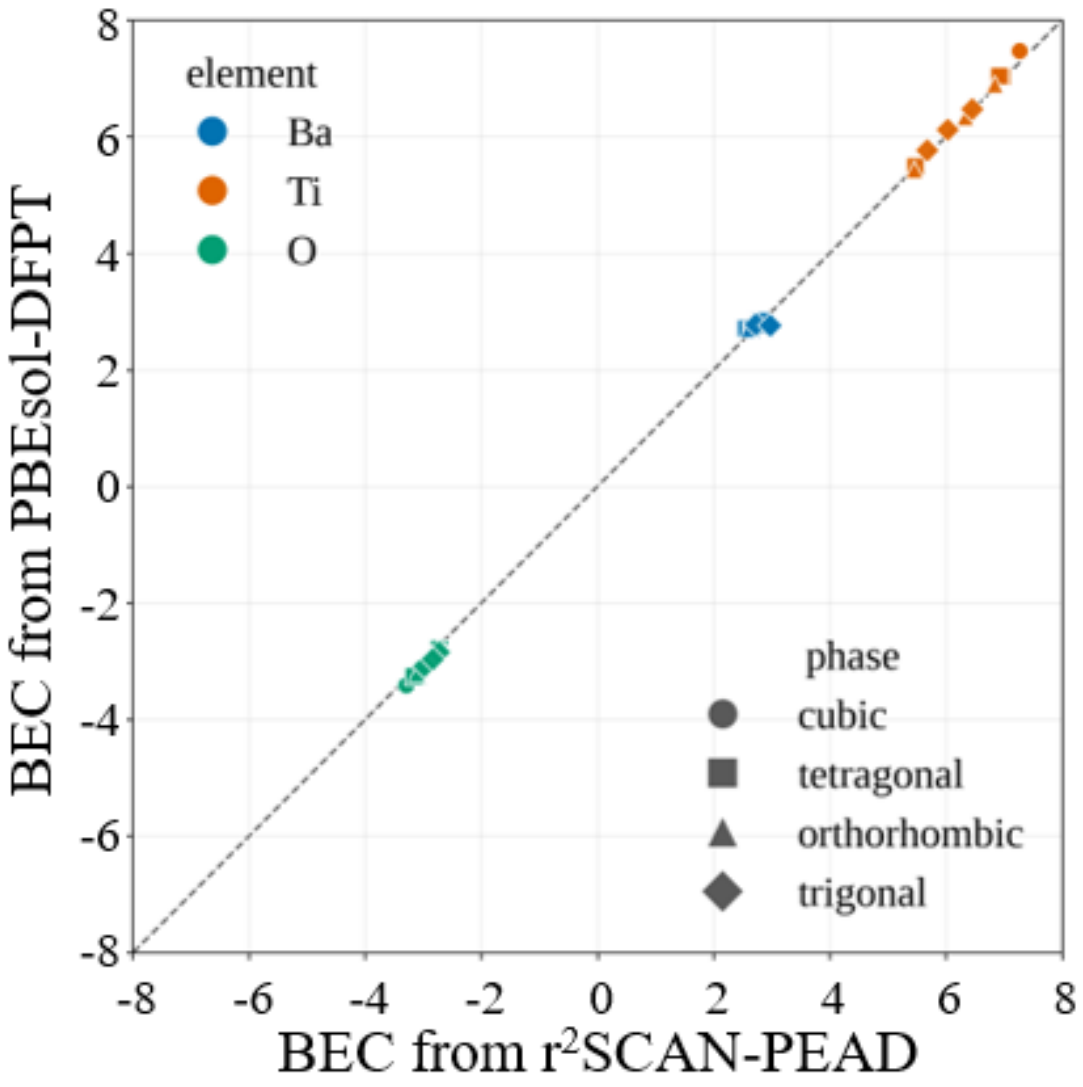

FIG. S2. Validation of the BECs used for the r$^2$SCAN-based MACEField model. Comparison of the diagonal BEC tensor components calculated using PBEsol DFPT and those independently obtained using the finite-electric-field PEAD method with the r$^2$SCAN functional. The dashed line indicates $y = x$. The mean absolute error between the two methods is 0.100 |e|.

As shown in Fig. S2, the diagonal BEC components obtained using the two approaches exhibit close agreement. The MAE between the PBEsol-DFPT and r$^2$SCAN-PEAD diagonal components is 0.100 |e|, supporting the use of PBEsol-DFPT BECs as training targets for structures generated at the r$^2$SCAN level.

The quantitative prediction errors of the resulting r$^2$SCAN-based MACEField model are summarized in Table SII. The MAEs in energy, atomic forces, and BECs are 0.17 meV atom$^{-1}$, 5.00 meV Å$^{-1}$, and 0.0039 |e|, respectively. These values demonstrate that the model accurately reproduces both the interatomic energetics and the local polarization response represented in the reference dataset.

The ability of the model to describe lattice dynamics was further examined by comparing phonon dispersions predicted by MACEField with those obtained from first-principles calculations. Figure S3 shows the phonon dispersions of the R, O, T, and C phases. The r$^2$SCAN-based MACEField reproduces the first-principles dispersions over the full Brillouin zone, with phonon-frequency MAEs ranging from 0.13 to 0.22 THz depending on the crystal phase. This agreement confirms that the model reproduces the local curvature of the potential-energy surface around

*Contact author: teru@iis.u-tokyo.ac.jp

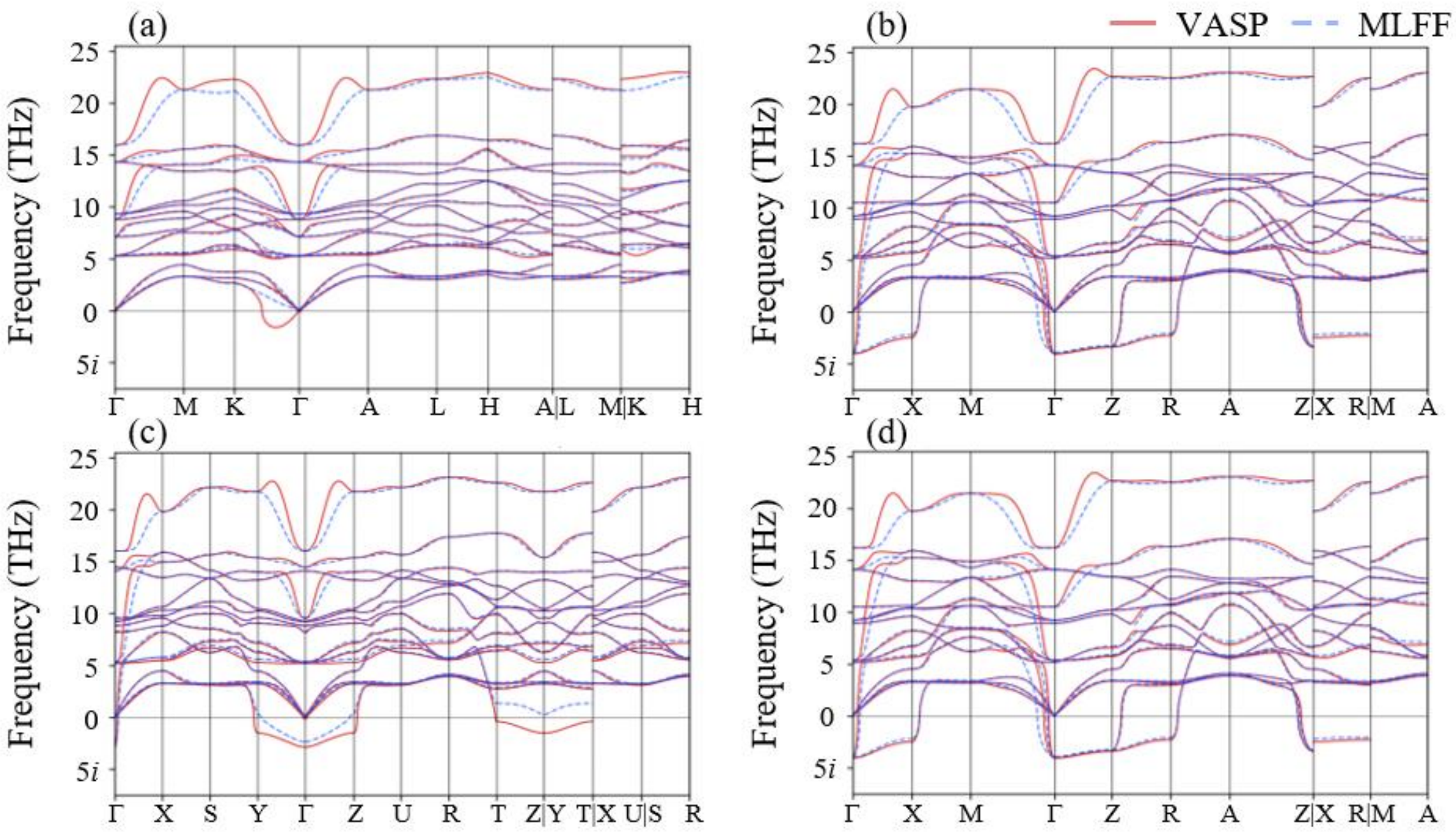

FIG. S3. Phonon dispersions of BTO in the (a) rhombohedral, (b) orthorhombic, (c) tetragonal, and (d) cubic phases. First-principles VASP results are shown by red solid lines, and predictions from the r$^2$SCAN-based MACEField model are shown by blue dashed lines. The model reproduces the first-principles phonon dispersions across all four phases, with phonon-frequency mean absolute errors ranging from 0.13 to 0.22 THz.

all four phases relevant to the finite-temperature behavior of BTO.

TABLE SII. Prediction errors of the MACEField models constructed using different exchange-correlation functionals. The MAEs in energy, atomic forces, and BECs are evaluated using the test dataset. The MAEs in phonon frequency are additionally evaluated for the R, O, T, and C phases of BTO.

| Model | | LDA | PBEsol | r$^2$SCAN |
|---|---|---|---|---|
| Energy MAE (meV atom$^{-1}$) | | 0.21 | 0.43 [2] | 0.17 |
| Force MAE (meV Å$^{-1}$) | | 4.71 | 14.9 [2] | 5.00 |
| BEC MAE (\|e\|) | | 0.0043 | 0.01 [2] | 0.0039 |
| Phonon MAE (THz) | R | 0.12 | 0.09 | 0.13 |
| | O | 0.08 | 0.06 | 0.15 |
| | T | 0.12 | 0.08 | 0.13 |
| | C | 0.20 | 0.15 | 0.22 |

## SIII. Identification of the Phase-Transition Temperatures

The phase-transition temperatures were determined from both the lattice parameters and directional changes of the local Ti–O off-centering during continuous-heating MD simulations. These directional angles, used only for phase-transition identification, are distinct from the polar angle relative to [001] analyzed in Fig. 2(a) of the main text.

Figure S4 shows a representative analysis for an $8 \times 8 \times 8$ supercell described by the r$^2$SCAN-based MACEField model. The reduced lattice parameters exhibit characteristic splitting and degeneracy associated with the R–O, O–T, and T–C phase transitions. Simultaneously, the mean directional angles of the Ti–O off-centering change across the corresponding temperature ranges.

To identify the transitions systematically, a Gaussian filter was applied to the temperature dependence of these directional angles, followed by evaluation of their temperature derivatives. Peaks in the derivatives were treated as transition candidates. Final transition temperatures were assigned only when anomalies were consistently observed in both the reduced lattice parameters and multiple orientational components, rather than from a single derivative extremum.

The black vertical dotted lines in Fig. S4 indicate the transition temperatures identified using this procedure. For the $8 \times 8 \times 8$ supercell described by the r$^2$SCAN-based model, the R–O, O–T, and T–C transition temperatures are 162.5, 240, and 335 K, respectively. These values represent apparent transition temperatures under the continuous-heating protocol employed here. The corresponding functional and supercell-size dependence is summarized in Fig. S1 and Table SI.

*Contact author: teru@iis.u-tokyo.ac.jp

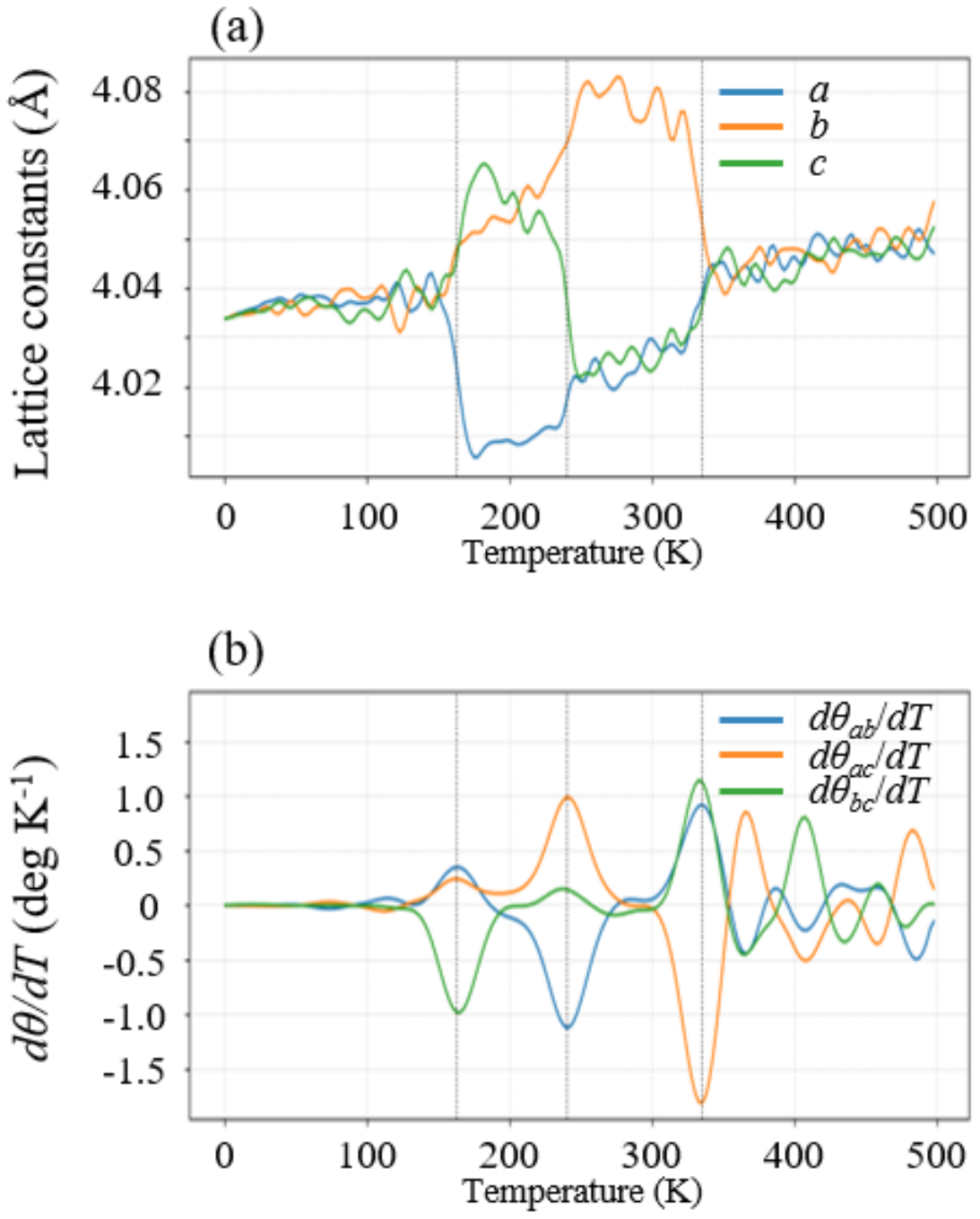

FIG. S4. Identification of the phase-transition temperatures of BTO during continuous-heating MD for an $8 \times 8 \times 8$ supercell described by the r²SCAN-based MACEField model. (a) Temperature dependence of the reduced lattice parameters $a$ (blue), $b$ (orange), and $c$ (green). (b) Temperature derivatives of the Gaussian-filtered directional angles of the Ti–O off-centering vector relative to the $ab$, $ac$, and $bc$ crystallographic planes. The blue, orange, and green curves represent $d\theta_{ab}/dT$, $d\theta_{ac}/dT$, and $d\theta_{bc}/dT$, respectively. Black vertical dotted lines indicate the R–O, O–T, and T–C transition temperatures determined from anomalies consistently observed in both the lattice parameters and the orientational derivatives. For this supercell, the transition temperatures are 162.5, 240, and 335 K, respectively.

## SIV. Ferroelectric Hysteresis and Polarization-Switching Dynamics

The response of the r²SCAN-based model to a large external electric field was further examined by calculating the polarization–electric-field hysteresis of tetragonal BTO at 270 K.

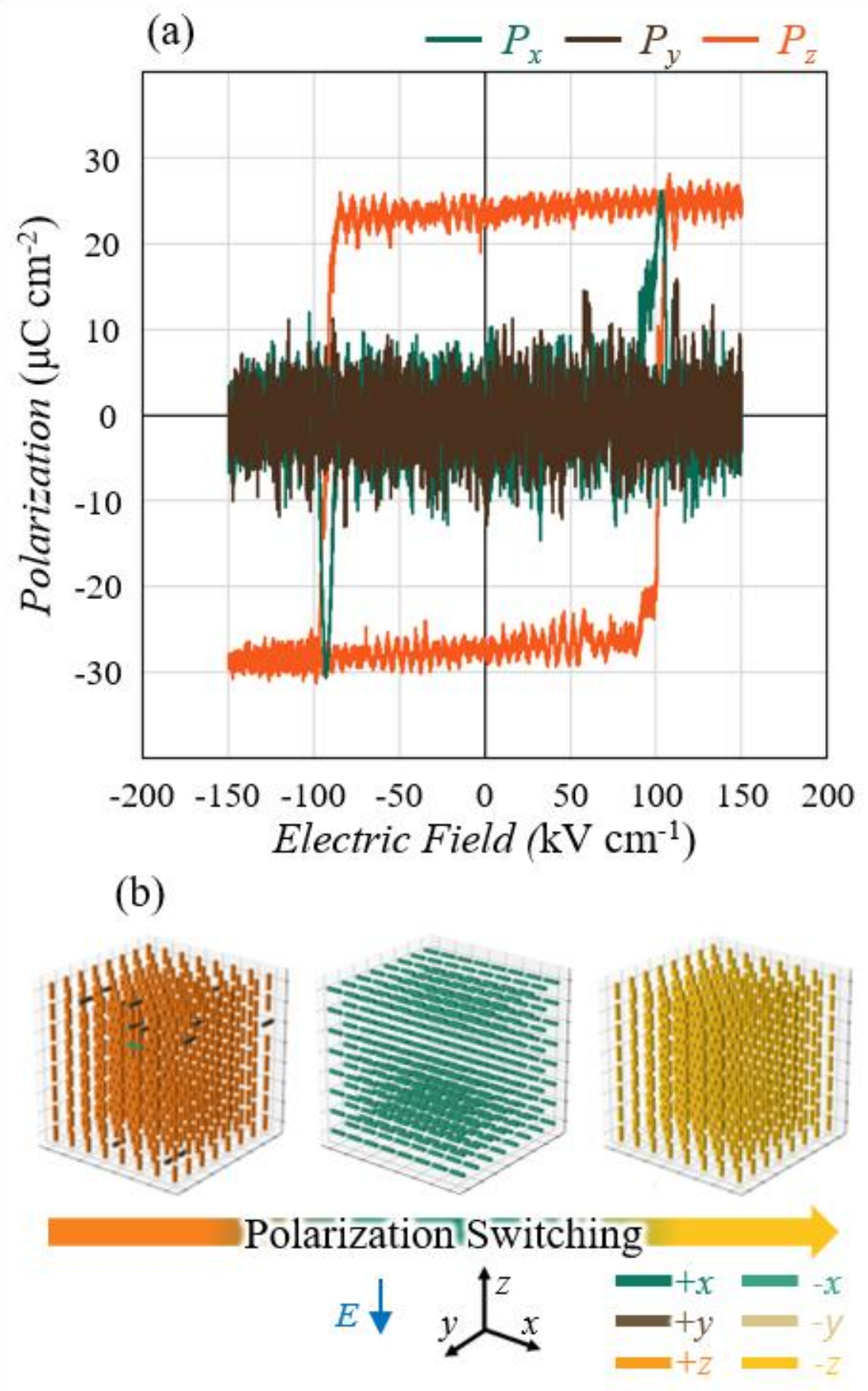

FIG. S5. Ferroelectric polarization switching in tetragonal BTO at 270 K simulated using the r²SCAN-based MACEField model. (a) Polarization–electric-field hysteresis loops obtained from the polarization components along the Cartesian $x$, $y$, and $z$ directions, which correspond to the crystallographic $a$, $b$, and $c$ directions, respectively. $P_x$, $P_y$, and $P_z$ are shown in green, brown, and orange, respectively. (b) Evolution of the populations of the six ⟨100⟩ orientations of the local Ti–O off-centering vectors during polarization switching. The orientations $\pm x$, $\pm y$, and $\pm z$ are color-coded as indicated in the legend. Before complete reversal of the primary polarization, the local Ti–O polar distortions transiently reorient toward an approximately 90° direction, accompanied by an increase in the transverse polarization component in (a).

Figure S5(a) shows the polarization components along the three crystallographic directions during the electric-field cycle. A characteristic ferroelectric hysteresis loop is obtained. The calculated remanent polarization is approximately 26 $\mu\mathrm{C\,cm^{-2}}$, in reasonable agreement with the experimentally reported value of approximately 21 $\mu\mathrm{C\,cm^{-2}}$ for BTO single crystals [5]. The calculated coercive field was

*Contact author: teru@iis.u-tokyo.ac.jp

99 kV cm$^{-1}$, substantially larger than the experimental value of 3.5 kV cm$^{-1}$ [5]. This discrepancy is expected. The dominant factor is the field-sweep rate: the present MD cycle spans 600 ps, some ten orders of magnitude faster than a laboratory measurement, and for nucleation-limited reversal the coercive field increases steeply with sweep rate. A second factor is the ideal periodic crystal considered here, which contains no pre-existing defects, surfaces, or other local inhomogeneities that facilitate polarization-reversal nucleation in real crystals. Recent atomistic simulations of BTO have likewise demonstrated that polarization reversal can involve nucleation and domain-wall formation, with the accessible switching pathway depending on temperature, mechanical conditions, and simulation conditions [6,7]. The absence of such nucleation-facilitating features may therefore contribute to the larger coercive field obtained in the present simulations.

To examine the corresponding atomistic switching mechanism, the populations of the local Ti–O off-centering vectors along the six crystallographic directions were tracked throughout the polarization-reversal process. As shown in Fig. S5(b), the local polar distortions do not reverse directly from one polarization direction to its opposite. Instead, a substantial population transiently rotates toward an approximately 90° orientation before reaching the final reversed state. This intermediate reorientation is accompanied by a transient increase in the transverse component of the macroscopic polarization in Fig. S5(a).

The approximately 90°-mediated switching pathway is consistent with switching behavior reported in experimental studies of BTO single crystals and with theoretical analyses based on Landau–Ginzburg–Devonshire theory [5,8]. This agreement provides an additional validation that the r$^2$SCAN-based MACEField captures not only macroscopic electric-field response but also the underlying local Ti–O polarization dynamics.

## SV. Robustness under Constrained In-Plane Lattice Conditions

In the strain-dependent calculations discussed in the main text, the in-plane lattice axes were constrained to maintain the prescribed biaxial strain. Because tetragonal temperature-dependent calculations used to establish the $\varepsilon_r$–$R_x$ relationship in the main text were performed without this constraint, we additionally evaluated the temperature dependence under the same in-plane lattice constraint used for the strain-dependent calculations. This comparison allows us to assess whether the different lattice constraints affect the temperature-dependent dielectric response or the relationship between $\varepsilon_r$ and $R_x$.

*Contact author: teru@iis.u-tokyo.ac.jp

Figure S6(a) shows the relative dielectric constant as a function of normalized temperature under the constrained in-plane lattice condition. The dielectric constant decreases from 2201 at 260 K to 731 at 320 K, demonstrating that the characteristic decrease in dielectric response with increasing temperature is retained even when the in-plane lattice degrees of freedom are constrained.

We further evaluated the field-induced orientational response $R_x$ under the same conditions. As shown in Fig. S6(b), the constrained-temperature dataset also exhibits a strong correlation between $\varepsilon_r$ and $R_x$. A linear regression gives $\varepsilon_r = 4.22 \times 10^4 R_x + 179$ with $R^2 = 0.907$.

The slope, $4.22 \times 10^4$, is within 5% of the value $4.42 \times 10^4$ obtained for the combined temperature- and strain-dependent relationship in the main text. This result confirms that the relationship between the dielectric response and the field-induced angular response of the local Ti–O polar structure is retained even when the in-plane lattice degrees of freedom are constrained.

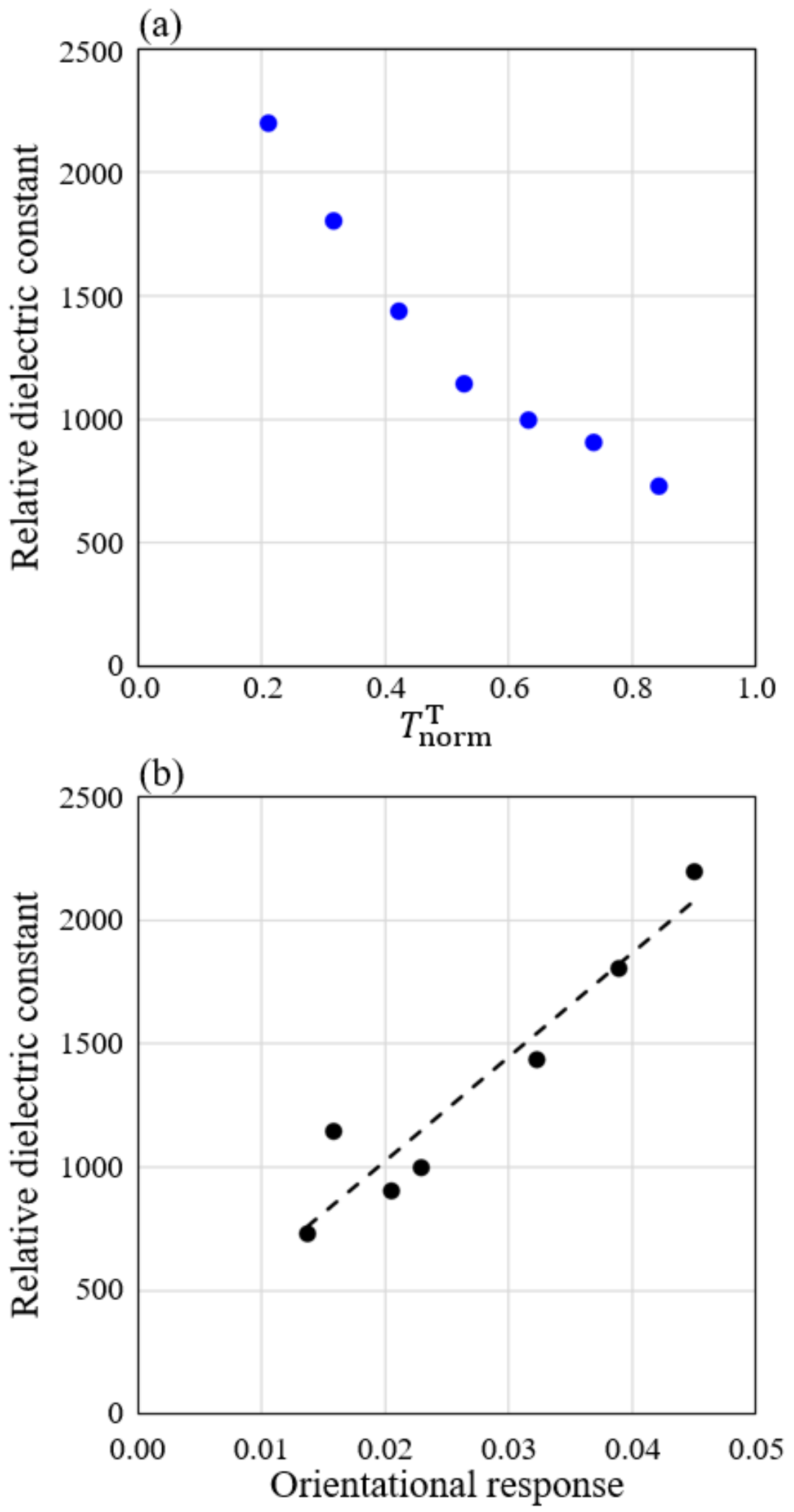

FIG. S6. **Temperature-dependent dielectric and orientational responses under constrained in-plane lattice conditions.** (a) Temperature dependence of the relative dielectric constant under the constrained in-plane lattice condition used for the strain-dependent calculations. The characteristic decrease in dielectric response with increasing temperature within the tetragonal phase is retained under the lattice constraint. (b) Relationship between $\varepsilon_r$ and the field-induced orientational response $R_x$ for the constrained-temperature dataset. For comparison, the unconstrained temperature-dependent and in-plane-strain-dependent datasets presented in the main text are also shown. The dashed line represents a linear regression to the constrained-temperature dataset, $\varepsilon_r = 4.22 \times 10^4 R_x + 179$ with $R^2 = 0.907$.

## SVI. Bond-Orientation-Resolved Oxygen Contribution to the Dielectric Response

The element-resolved analysis in Fig. 1(b) of the main text shows that O provides a major contribution to the dielectric response. To determine whether this response is distributed equally among the three Ti–O bond orientations, we further decomposed the O contribution according to the direction of the corresponding Ti–O bond. O1, O2, and O3 denote oxygen sites whose Ti–O bonds are oriented along the crystallographic $c$, $b$, and $a$ directions, respectively. Because the electric field is applied along the crystallographic $a$ direction, O3 corresponds to the oxygen site bonded parallel to the applied field.

Figure S7(a) shows the fraction of the total O contribution arising from each oxygen orientation as a function of normalized temperature $T_{\mathrm{norm}}^{T}$. The contributions from O1, O2, and O3 sum to 100% at each temperature. O3 consistently provides the dominant contribution, accounting for approximately 67–85% of the O response over the examined temperature range.

This site selectivity is consistent with the strong anisotropy of the oxygen BEC. As shown in Fig. S7(b), the O3 site exhibits a large-magnitude BEC component along the applied-field direction, $Z_{xx}^{*} \approx -5.1$ to $-5.2$, whereas the corresponding components for O1 and O2 are approximately $-2.0$. Such strong anisotropy of the oxygen BEC is well established in BTO. Previous first-principles calculations for cubic BTO at 0 K reported oxygen BEC components of approximately −5.71 |e| parallel to the Ti–O bond and −2.15 |e| perpendicular to it [9]. The finite-temperature values obtained here are consistent with this anisotropy, the somewhat smaller longitudinal magnitude being expected for thermally averaged configurations. Thus, the polarization response along the $a$ direction is strongly weighted toward the oxygen site whose Ti–O bond is parallel to the applied field. Because the dielectric contribution also depends on the field-induced atomic displacements, the BEC anisotropy is consistent with, rather than solely responsible for, the dominant O3 response. In the cubic phase, O1, O2, and O3 are distinguished here by their Ti–O bond orientations relative to the applied field rather than by crystallographic inequivalence.

*Contact author: teru@iis.u-tokyo.ac.jp

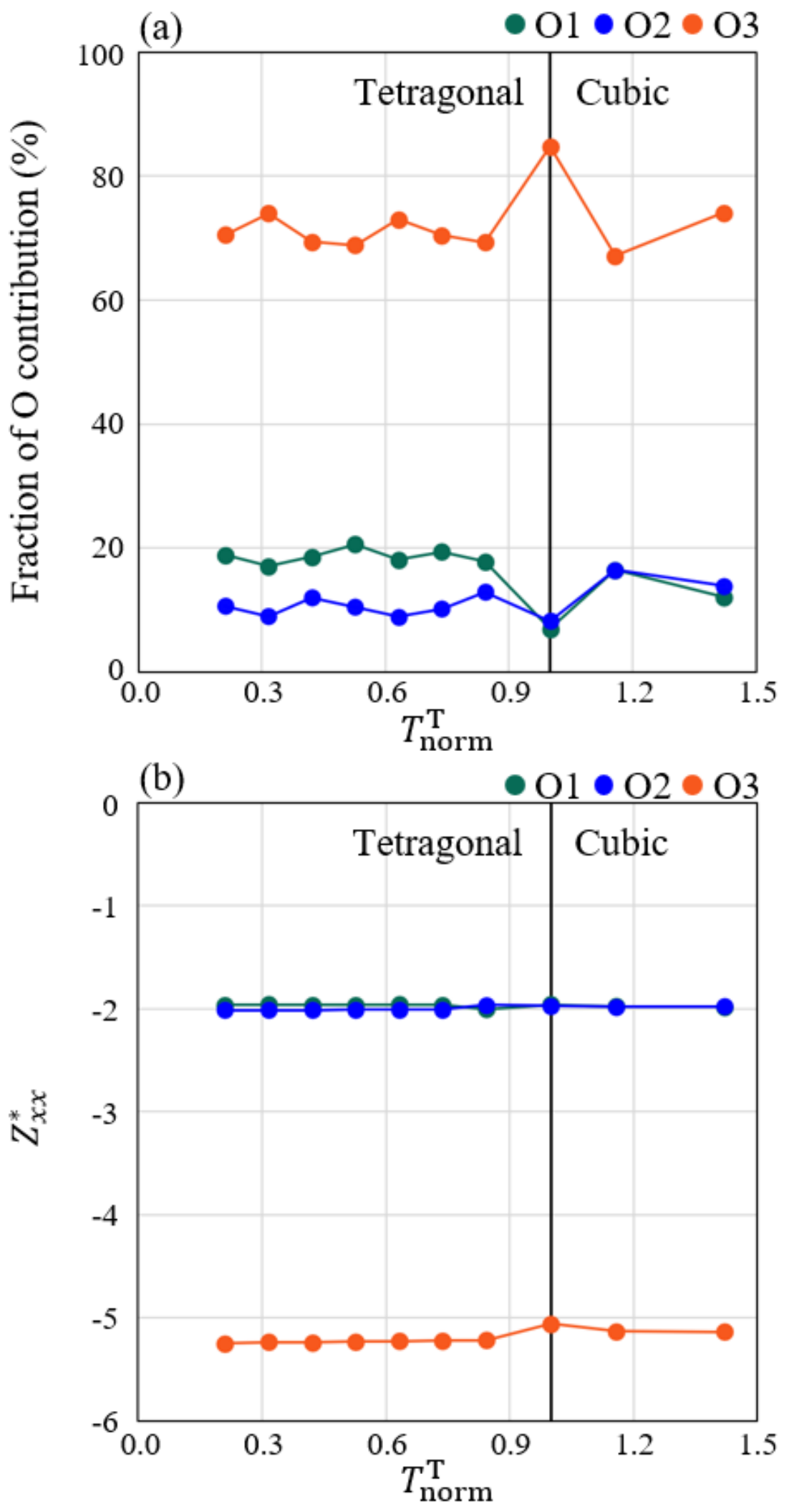

FIG. S7. **Bond-orientation-resolved oxygen contributions to the dielectric response.** O1, O2, and O3 denote oxygen sites whose Ti–O bonds are oriented along the crystallographic $c$, $b$, and $a$ directions, respectively, while the electric field is applied along the crystallographic $a$ direction. (a) Fraction of the total O contribution arising from O1, O2, and O3 as a function of normalized temperature $T_{\mathrm{norm}}^{T}$. The three contributions sum to 100% at each temperature. The $a$-bonded O3 site provides the dominant contribution throughout the examined temperature range. (b) Corresponding $Z_{xx}^{*}$ components of the three O sites as a function of $T_{\mathrm{norm}}^{T}$. O3 exhibits the large longitudinal BEC component along the applied-field direction, whereas O1 and O2 exhibit smaller transverse components. In both panels, O1, O2, and O3 are shown in green, blue, and orange, respectively, and the black vertical solid line marks the T–C phase transition at $T_{\mathrm{norm}}^{T} = 1$.

## SVII. Extension of the Local-Response Analysis Across the T–C Transition

The local-response analysis in the main text focuses on the tetragonal phase in order to identify the microscopic origin of the temperature-dependent dielectric response without crossing a structural phase boundary. We additionally examined the Ti–O local response at and above the T–C transition to determine whether the same microscopic picture extends into the cubic phase.

Figure S8(a) compares the mean local Ti–O off-centering magnitude, $\langle | \boldsymbol{u} | \rangle$, with the magnitude of the vector-averaged off-centering, $| \langle \boldsymbol{u} \rangle |$. At 350 and 375 K, $\langle | \boldsymbol{u} | \rangle$ is 0.1750 and 0.1755 Å, respectively, whereas $| \langle \boldsymbol{u} \rangle |$ decreases to 0.0117 and 0.00727 Å, corresponding to only 6.68% and 4.14% of the respective local magnitudes. Thus, individual Ti–O units remain locally off-centered in the cubic phase, while their differently oriented off-centering vectors largely cancel upon averaging. At 335 K, close to the calculated T–C transition, the corresponding ratio remains larger at 27.0%.

We next examined whether the same orientational descriptor that organizes the dielectric response within the tetragonal phase remains applicable across the T–C transition. Figure S8(b) compares the tetragonal temperature-dependent data shown in green with the data at and above the T–C transition shown in blue. Both datasets exhibit an approximately linear dependence of $\varepsilon_r$ on $R_x$, as indicated by the corresponding dashed fits. The data at and above the transition are described by $\varepsilon_r = 6.63 \times 10^4 R_x - 2.24 \times 10^3$ with $R^2 = 1.00$.

Together with the tetragonal results, this shows that $R_x$ consistently tracks the dielectric response on both sides of the T–C transition. Within the tetragonal phase the temperature- and strain-dependent data follow a single linear relation. Across the T–C transition both the slope and the intercept change, so the quantitative mapping is not universal; nevertheless the monotonic correlation between $R_x$ and $\varepsilon_r$ persists, showing that the same field-induced orientational response remains a relevant local descriptor on both sides of the transition.

*Contact author: teru@iis.u-tokyo.ac.jp

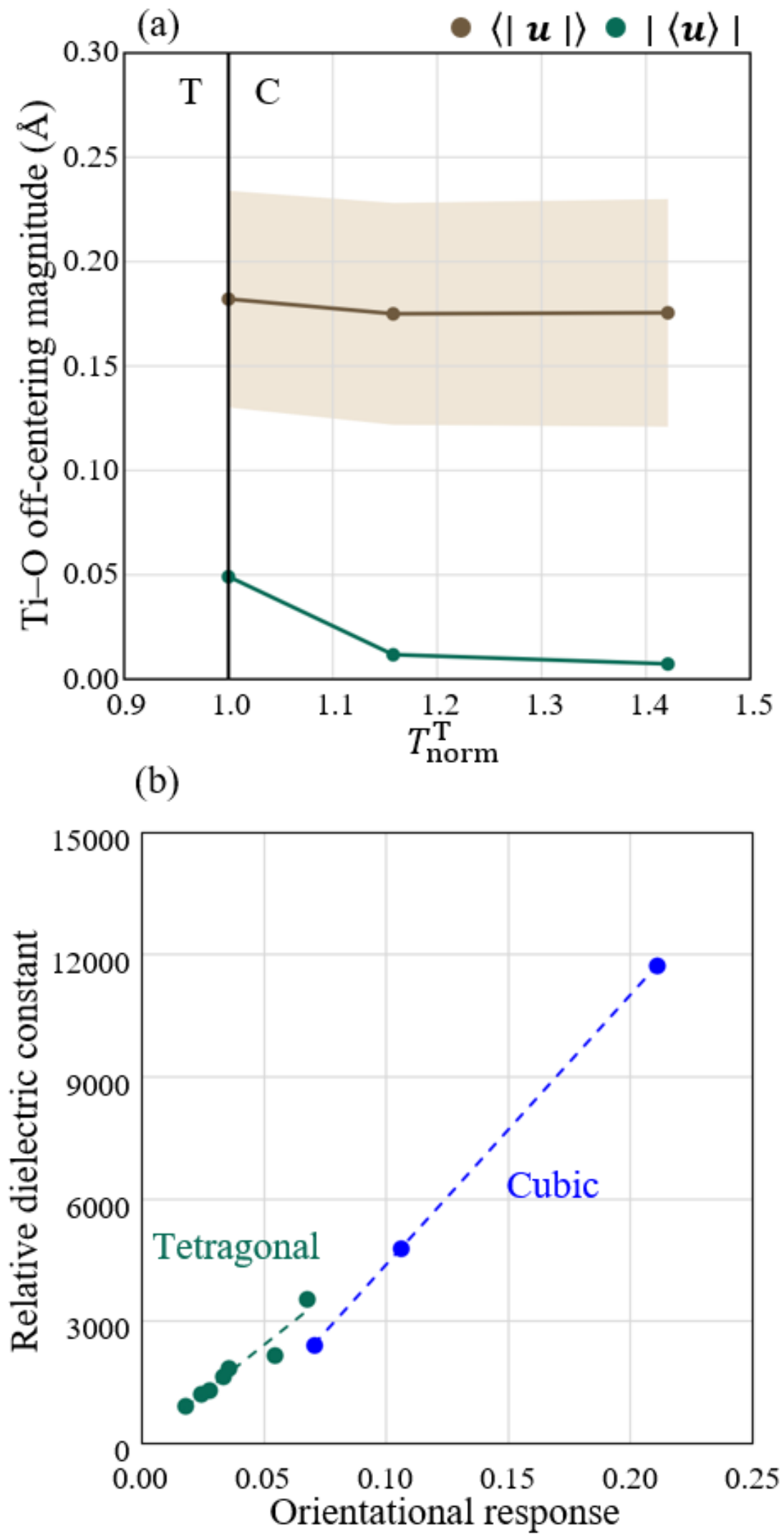

FIG. S8. **Extension of the local-response picture across the T–C transition.** (a) Mean local Ti–O off-centering magnitude, $\langle | \boldsymbol{u} | \rangle$, and magnitude of the vector-averaged off-centering, $| \langle \boldsymbol{u} \rangle |$, as a function of normalized temperature $T^{T}_{\mathrm{norm}}$. Brown symbols show $\langle | \boldsymbol{u} | \rangle$ with standard deviations, and green symbols show $| \langle \boldsymbol{u} \rangle |$. The local off-centering remains finite above the transition, whereas its vector average becomes much smaller. The black vertical solid line marks the T–C transition at $T^{T}_{\mathrm{norm}} = 1$. (b) Relative dielectric constant $\varepsilon_r$ plotted against the orientational response $R_x$. Tetragonal and at/above-transition data are shown in green and blue, respectively; dashed lines of the corresponding colors indicate linear fits. Both regimes show a clear $\varepsilon_r$–$R_x$ correlation.

## REFERENCES

[1] R. Sahashi, P.-Y. Chen, and T. Mizoguchi, LDA-based machine learning force field for accurate electric-field-driven ferroelectric response in $BaTiO_3$, J. Ceram. Soc. Jpn. **134**, 431 (2026).

*Contact author: teru@iis.u-tokyo.ac.jp

[2] P.-Y. Chen and T. Mizoguchi, Transition from homogeneous to domain-wall-mediated polarization switching in $BaTiO_3$: A machine-learning molecular dynamics study, Phys. Rev. B (to be published).
[3] W. J. Merz, The electric and optical behavior of $BaTiO_3$ single-domain crystals, Phys. Rev. **76**, 1221 (1949).
[4] L. Gigli, M. Veit, M. Kotiuga, G. Pizzi, N. Marzari, and M. Ceriotti, Thermodynamics and dielectric response of $BaTiO_3$ by data-driven modeling, npj Comput. Mater. **8**, 209 (2022).
[5] B. Jiang, Y. Bai, W. Chu, Y. Su, and L. Qiao, Direct observation of two 90° steps of 180° domain switching in $BaTiO_3$ single crystal under an antiparallel electric field, Appl. Phys. Lett. **93**, 152905 (2008).
[6] H. Azuma, T. Ogawa, S. Ogata, R. Kobayashi, M. Uranagase, T. Tsuzuki, and F. Wendler, Unique temperature-dependence of polarization switching paths in ferroelectric $BaTiO_3$: A molecular dynamics simulation study, Acta Mater. **296**, 121216 (2025).
[7] P.-Y. Chen and T. Mizoguchi, Effect of uniaxial compressive stress on polarization switching and domain wall formation in tetragonal phase $BaTiO_3$ via machine learning potential, Mater. Des. **265**, 115851 (2026).
[8] Y. Li, J. Wang, and F. Li, Intrinsic polarization switching in $BaTiO_3$ crystal under uniaxial electromechanical loading, Phys. Rev. B **94**, 184108 (2016).
[9] Ph. Ghosez, J.-P. Michenaud, and X. Gonze, Dynamical atomic charges: The case of $ABO_3$ compounds, Phys. Rev. B **58**, 6224 (1998).

*Contact author: teru@iis.u-tokyo.ac.jp